\documentstyle[12pt]{article}
\begin{document}
\begin{titlepage}
\begin{flushright}
          \begin{minipage}[t]{12em}
          \large UAB--FT--380\\
                 hep-astro/9601145\\
                 January 1995
          \end{minipage}
\end{flushright}
\vspace{\fill}
\begin{center}
\baselineskip=2.5em
{\huge Baryonic Dark Matter: Theory and Experiment. Overview.
$^{\dag}$}
\end{center}
\vspace{\fill}
\begin{center}
{\sc Eduard Mass\'{o}}\\
     Grup de F\'\i sica Te\`orica and Institut de F\'\i sica 
     d'Altes Energies\\
     Universitat Aut\`onoma de Barcelona\\
     08193 Bellaterra, Barcelona, Spain
\end{center}
\vspace{\fill}
\begin{center}
\large ABSTRACT
\end{center}
\begin{center}
\begin{minipage}[t]{36em}
The general arguments for baryonic and galactic dark matter are
presented. Limits coming from a variety of theoretical
considerations and observations are discussed. The surviving
candidates for galactic baryonic dark matter seem most likely to be
in the form of compact objects and could be in one of two mass
windows: either in the brown dwarf regime or in the mass range
corresponding to supermassive black holes. Microlensing towards LMC
is probing the first window. It is important to keep in mind that
these experiments may detect compact heavy objects, independent of
their constituency.
\end{minipage}
\end{center}
\vspace{\fill}
{\noindent\makebox[10cm]{\hrulefill}\\
\footnotesize
\makebox[1cm][r]{$^{\dag}$} 
Written version of a talk at IV Workshop on Theoretical and 
Phenomenological Aspects of Underground Physics,
TAUP95, Toledo (Spain), Sept. 17-21. To be published in 
the Proceedings, edited by M. Fat\'{a}s.}
\end{titlepage}
\clearpage
\vbox{\vspace{3em}}

\section{INTRODUCTION}

Dark matter (DM) has been proposed in different contexts and scales.
There are very strong observational indications of its existence in
galactic halos and in clusters. DM associated to our galactic disk
has been proposed but its existence is dubious. In inflationary
models one has a cosmological background of DM. 

As we will see, there are strong indications that there is
baryonic DM. A priori, this does not preclude the existence or even
dominance of non-baryonic (mainly elementary particle) DM. 

Of course in both cases the
existence of dark matter is inferred from the comparison of the
corresponding halo or baryonic density with the visible matter
density. 

In this workshop non-baryonic DM was reviewed by Berezinsky 
\cite{berezinsky}. In my talk I will concentrate on baryonic and
galactic dark matter, which are sometimes thought as the ones for
which we have the strongest evidence. Other reviews on baryonic DM 
are cited in \cite{carr}. 

\section{LIGHT AND MASS}
 
We start quoting a recent estimation by Persic and Salucci
\cite{persic} of the density of visible matter in the Universe
     
\begin{equation} \label{rhov1}
 \rho_v = \sum \int dL\ \phi(L) L \Upsilon_v  \ \ . 
\end{equation}
  
The integration is over the luminosity $L$ and
the sum extends to different galaxy types and hot gas in clusters
and groups of galaxies. For each case, Persic and Salucci take the
corresponding luminosity function $\phi(L)$ and mass-to-light ratio
$\Upsilon_v= M_v/L$. Their careful scrutiny leads to
   
\begin{equation} \label{rhov2}
 \Omega_v = \frac{\rho_{v}}{\rho_{crit}} \simeq 0.003  \ \ . 
\end{equation}

As usual we normalize to the critical density  $\rho_{crit}= 0.69
\times 10^{11} h_{50}^2\ M_\odot/Mpc^3$; $H_0 = 50\ h_{50}\
km/s/Mpc$. 

The key question is: Can we extrapolate this low density to other
regions of the Universe?  In order to discuss whether in general
light is a good tracer of mass  in the Universe, one considers the
mass-to-light ratio
  
\begin{equation} \label{def1}
 \Upsilon = \frac{M}{L}   \ \ ,
\end{equation}
conventionally normalized to the value $\Upsilon_\odot =
M_\odot/L_\odot$. In the cores of galaxies, the visible contribution
to $\Upsilon$ is a few times the value of $\Upsilon_\odot$. If light
were a good tracer of mass, the value of $\Upsilon$ would be about
the same when going to different astronomical systems. However, the
values found in the cores of galaxies cannot be extrapolated to
regions of lower luminosity. One indeed observes much larger values
of $\Upsilon$ in a variety of systems with scales larger than 
galaxies. This shows that there is non-visible matter in the
Universe; the DM manifesting itself through gravitational
effects. One of the main goals of modern physics is to discover the
nature of this DM. 

From two points of view, the DM problem should be regarded as not
too surprising. First, mass and light are already not well
correlated even in main-sequence stars in our solar neighbourhood:
the quantity $\Upsilon_{star}=M_{star}/L_{star}$ is strongly
varying with $M_{star}$. It is the low-luminosity stars that
contribute the most to mass: stars with $L<L_\odot$ contribute at
least 75\% of the total mass
\cite{binney}. Second, there are many forms of stable matter that we
may conceive of, some visible and some dark. From this point of
view, it would be naive to think that all matter in the Universe
should only be in visible form. Here we are thinking of very
conventional forms of matter like brown dwarves but also of neutral
stable particles arising in extensions of the Standard Model of
particle physics, like axions or neutralinos.

\section{BARYONIC DARK MATTER}

We will discuss three important issues in connection with baryonic DM: 

(a) The flatness of galactic  rotation curves, which is the
strongest observational evidence of DM. Galactic dark halos are
the most natural place for dark baryons.

(b) Big bang nucleosynthesis, since it is sensitive to the baryonic
density, which includes visible and dark baryons.

(c) The detected (LMC) microlensing events, that probably show that
there is some baryonic DM in our Galaxy. However, it is not clear
whether it is halo DM.

In this Section we shall discuss items (a) and (b). The
microlensing events (c) will be discussed along with other limits on
astrophysical DM in Sect.4.

Apart from all that, we would like to mention the recent
ROSAT X-ray observations that seem to indicate that clusters and
groups of galaxies could contain a relatively large density of gas
\cite{white}, leading to a contribution $\Omega_B \sim 0.01$.
  
\subsection{Galactic rotation curves} 

The rotation curves $v(r)$ of spiral galaxies are well measured for
distances $r$ to the galactic center larger than the disk radius,
$r_{vis}$. Stars, gas and dust are inside $r_{vis}$, and
contribute a mass $M_{vis}$. If all the mass of the galaxy were in
such a visible form we would have that, for $r$ much larger than
$r_{vis}$,

\begin{equation}\label{newton}
     \frac{v^2(r)}{r} \simeq \frac{G M_{vis}}{r^2} \ \ .   
\end{equation}
 
Thus, we would expect a keplerian fall-off of the rotation curves for
large $r$
  
\begin{equation} \label{kepler}
      v(r) \simeq \left( \frac{G M_{vis}}{r} \right) ^{1/2} \sim
            \frac{1}{r^{1/2}}  \ \ .
\end{equation}  
 
Instead of this behavior, it is measured that
 
\begin{equation}
      v(r) \sim constant  \ \ ,
\end{equation} 
for large $r$. Flat rotation curves are the common trend of all
spiral galaxies that have been studied \cite{rotationcurves}.

It is accepted that the flat rotation curves of galaxies are
the gravitational effect of galactic dark halos. It is easy to
see which is the required behavior of the dark mass density
at large $r$. Assume spherical symmetry and consider a sphere of
radius $r$ and with center the galactic center. In a first
approximation it is clear that the mass contained in such
sphere, $M(r)$, should increase linearly with the radius, i.e. $M(r)
\sim r$, since then we will have the observed behavior
 
\begin{equation}  \label{dark}
      v(r) \simeq \left( \frac{G M(r)}{r} \right) ^{1/2} \sim 
constant  \ \ .
\end{equation}
 
This simple analysis tells us that for large $r$ the dark
mass density should behave approximately as
 
\begin{equation}  \label{rho}
\rho(r) \sim \frac{M(r)}{r^3} \sim \frac{1}{r^2}  \ \  .
\end{equation}

More precise determinations of the mass distribution of the dark halo
of our Galaxy use also the assumption of spherical symmetry. (This
assumption is supported by numerical calculations, but it should be
kept in mind that it is not observationally proved.) The dark halo
density is then taken as a (non-singular) generalization of
eq.(\ref{rho}) 

 \begin{equation}    \label{rhoDM}
      \rho(r) = \rho(r_0) \, \frac{a^2+r_0^2}{a^2+r^2} \ \ ,
\end{equation}
with $r_0 \simeq 8.5\ kpc$ the distance of our solar system to the
center of the Galaxy. This halo form is sometimes referred as the
standard halo. The parameters $a$ and
$\rho(r_0)$ have been determined by several authors
\cite{milkywaya,milkywayrho}. One finds a local density

\begin{equation}   \label{rhor0}
\rho(r_0) \simeq 0.01\ \frac{M_{\odot}}{pc^3}    \ \ ,
   \end{equation} 
and a core radius
 
\begin{equation}  \label{a}
a \approx 6\ kpc    \ \ .
\end{equation}
 
The local density is obviously crucial for experiments that aim at
detecting the galactic DM, but unfortunately the number
quoted in eq.(\ref{rhor0}) is subject to an uncertainty of about a
factor 2 \cite{milkywaya,milkywayrho}. Also, the core radius $a$ has
a large incertitude; one finds values from $a=2\ kpc$ to $a=8\
kpc$ in the literature. A related issue concerns the total extent
and mass of the Milky Way halo. Our galactic halo is probably more
than a factor 10 larger that the galaxy itself in both mass and size
\cite{fich}.

The estimation of the galactic halo density depends on this total
extent, and on the objects we choose to study the velocity
dispersion. All the estimations lead to values higher than 2 \%,

 \begin{equation}    \label{rhogal}
      \Omega_{halo}  >  0.02  \ \ ,
\end{equation}
and a fortiori higher than the visible density (\ref{rhov2}).

\subsection{Big bang nucleosynthesis}

Big bang nucleosynthesis (BBN) allows us to understand
the formation of light elements in the early Universe (see the talk
of G. Steigman at this workshop \cite{steigman}). Their relative
amounts can be calculated as a function of the number density of
baryons. The theoretical calculations have then to be compared to the
primordial abundances, obtained from the observed yields and taking
into account the chemical evolution of elements. The analysis leads
to a bound on the present density of baryons  $\rho_{B}$. Walker
{\it et al.} \cite{walker} have obtained  

\begin{equation}
\label{range}
 0.04 \leq \Omega_B h_{50}^2 \leq 0.06    \ \ .   
\end{equation}

In updated analyses, Olive and Scully \cite{olive} have
confirmed the range in eq.(\ref{range}) while Kernan and Krauss
\cite{kernan} find the same lower bound than in eq.(\ref{range}) but
a lower upper bound ($\Omega_B h_{50}^2 \leq 0.044$). 

Let us now compare the lower bound on the baryon density in
eq.(\ref{range}) with the visible density, eq.(\ref{rhov2}),
 
\begin{equation}
\label{Bv} \frac{\Omega_B}{\Omega_v} \geq 
\frac{13.}{h_{50}^2} \geq 3. \ \ ,    
\end{equation} 
where the last inequality follows from $h_{50} \leq 2$. The meaning
of eq.(\ref{Bv}) is that at least 75\% of the baryons in the
Universe are dark. More firm conclusions on this baryonic DM would be
obtained sharpening the value of the Hubble constant $H_0$; for
example, a value $H_0=50\ km/s/Mpc$ would imply that dark baryons
constitute more than 90\% of the total baryonic matter. 

BBN shows that the bulk of baryons in the Universe is dark, and
this is based on the lower bound in eq.(\ref{range}). In turn, the
precise value for this lower bound depends quite strongly on the
deuterium D and $^3$He abundances. Recently, there has been some
debate on these abundances. 

First, there has been a claim of BBN crisis \cite{hata,steigman} when
using specific models of chemical evolution for D. However, a very
recent reanalysis \cite{copi} of BBN conclude that predicted and
measured abundances of light elements are consistent with 95\%
credibility if $\Omega_B h_{50}^2 \geq 0.028$. We need more work
along these lines to see whether the standard scenario is in
trouble or not. 

Second, some reports \cite{carswell} on detection of D in
low-metallicity, high redshift quasar absorption systems could
indicate high primordial D densities which could affect the bounds
on the baryonic density of the Universe. Based on these measurements
(and on the X-ray observations we quoted in Sect.2), Dar \cite{dar}
claims most of the baryons in the Universe are visible. However,
there are other observations in absorption systems which lead to
results that do not coincide with Ref.\cite{carswell}, perhaps
showing that there are large systematic corrections to this type of
measurements. Again, more observations and theoretical work
would be welcome, since this important point should be
elucidated.       

A very interesting study has been performed by Fields and Olive
\cite{fields}. Their idea is that, since using D and $^3$He
abundances is controversial, one may use solely the more solid $^4$He
and $^7$Li yields in the nucleosynthesis analysis. They obtain 
(95\% CL)

\begin{equation} \label{fo}
 0.02 \leq \Omega_B h_{50}^2     \ \ .   
\end{equation}

As expected, the fact of not using D and $^3$He abundances as input
weakens the bound (\ref{fo}). Still, since $h_{50} \leq
2$, there are more dark than visible baryons.

Finally, we would like to comment that it seems difficult to evade
the bounds on the baryonic density. One can relax them a bit by
invoking non-standard scenarios like inhomogeneous nucleosynthesis
\cite{inho}, or late decaying particles \cite{decay}. 

\section{BARYONIC DARK MATTER CANDIDATES} 

In the last Section we saw that rotation curves are
compelling evidence for galactic DM, and we also saw that comparing
the baryonic density deduced from BBN we find evidence for dark
baryons. It turns out that: 

(A) the ranges for $\Omega_{halo}$ and $\Omega_{B}$ overlap. 

(B) the most natural place for dark baryons is in the Galaxy. 

The natural questions are: Are there dark baryons in the galactic
halo? What is the contribution of dark baryons to the halo
density? We will analyze in turn the plausibility of the different
candidates for baryonic DM, starting with diffuse matter. 

Dust forming the galactic DM would lead to too much starlight
extinction, and is therefore excluded. Another conservative
assumption on diffuse baryonic DM is that it is in the form of gas.
However, gas with the required mass density and  at the virial
temperature of the Galaxy,
 
\begin{equation} \label{tvirial}
 T_{virial} \sim m_p\ <v^2>\ \sim 2 \times 10^{6} K  \ \ ,    
\end{equation}
would emit soft X-rays that we should have detected.

Although this may exclude gas as DM, a word of caution is needed
here. There are galactic scenarios in which cold  molecular H$_2$
clouds exist at galactocentric distances larger than about 10 kpc
\cite{depaolis}. These clouds could contribute substantially to the
galactic halo density. In Ref.\cite{depaolis}, the authors propose
ways to observe such clouds. Another scenario has been recently
developed in Ref.\cite{gerhard}. Here, substantial quantities of
cold gas are stabilized my macho clusters.     

Next, we turn our attention to galactic compact objects. We will
review in the rest of the Section the constraints we have on their
presence in the galactic halo. We first discuss limits on small solid
objects, $M << O(M_{\odot})$. Planet-like objects with $M <
O(M_{\odot})$ and  the issue of microlensing and brown dwarves are
discussed in Subsection 4.2. The last Subsection is devoted to very
heavy galactic objects, 
$M > O(M_{\odot})$.
 
\subsection{Snowballs}
   
Could snowballs (formed by cold condensed hydrogen) be the baryonic
DM in our halo? There are strong constraints on such objects.

For very small masses, a galactic population of snowballs
has a high density (if they have to form the whole of the dark
halo), and collisions among them would  lead to destruction of such
objects. This establishes a lower bound \cite{hegyi}

\begin{equation}     \label{snowball1} 
M \geq 1\ g  \ \ .
\end{equation}

Another effect is that such solid objects would continuously hit the
Earth. The observed frequency of interstellar meteors and comets
exclude the range \cite{hills}

\begin{equation}     \label{snowball2} 
10^{-3}\ g \leq  M  \leq 10^{22}\ g \ \ .
\end{equation}

\subsection{Brown Dwarves, Machos and Microlensing}

The nuclear-ignition threshold for a hydrogenous compact object is

\begin{equation}     \label{nucl} 
 M_{nucl} = 0.08 M_{\odot} \ \ .
\end{equation}
 
Below this mass, the objects are called brown dwarves. 

Thermal evaporation of brown dwarves might be 
an important effect. If we require that a brown dwarf with mass $M$
has not evaporated in a galactic time-scale we can set an absolute
lower bound \cite{derujula1}
     
\begin{equation}     \label{evaporation} 
  M  \geq M_{evap} \simeq 10^{-7} M_{\odot} \ \ .
\end{equation}

Failed stars that have not yet evaporated are a galactic baryonic DM
candidate. Their masses can be in the brown dwarf window: 
\begin{equation}  \label{window1} 
10^{-7} M_{\odot} \simeq M_{evap}
\leq  M \leq
 M_{nucl} \simeq 10^{-1} M_{\odot}  
\end{equation}
 We will now examine
constraints on such objects.

Although brown dwarves do not start nuclear reactions in their
interior, they still generate some infrared luminosity. This  has
been investigated by Kerins and Carr \cite{kerins}. If brown dwarves
with a common mass $M$ form the galactic halo, the expected
distance $<d>$ to the nearest one is given by 

\begin{equation}     \label{d} 
<d>\  \simeq\  1.2\ M_{0.1}^{1/3}\ pc  \ \ .
\end{equation}
($M_{0.1}$ is the mass of the compact object in units of $0.1
M_{\odot}$.) For a normal star, a distance $O(pc)$ is small. However,
a brown dwarf is very cold and  then the expected flux \cite{kerins}
is too low to have been detected by IRAS. The European satellite ISO
will be sensitive to brown dwarves with $M>.01M_{\odot}$
\cite{kerins}. In any case, the microlensing effect is a  much more
efficient way to detect brown dwarves if they form a substantial
part of the galactic halo, so that we deserve the rest of the
Subsection to that effect.   

In 1986, Paczy\'nski \cite{pac} showed how to detect massive halo
objects in our galactic halo by means of the gravitational lensing
effect on the images of stars in the Large Magellanic Cloud (LMC).
The gravitational deflection of the light of the LMC star by the
halo object leads to a time-dependent magnification of the
original brightness of the source. The effect depends on the
gravitational field of the object, and not on the composition.
Microlensing events could thus be a signature of the presence of
dark galactic objects, independently of whether or not they
are hydrogenous (brown dwarves). We will refer to the deflector
as macho, for massive compact halo object.  

The scale of microlensing is set by the Einstein radius
\begin{eqnarray} \label{einstein}
R_E & = &  2\ \sqrt{GMDx(1-x)} \nonumber \\ 
    & = &  6.7 AU\ M^{1/2}_{0.1}\  
           \left(  \frac{x(1-x)D}{55\ kpc}  \right)^{1/2}   
\end{eqnarray}
Here $D=55\ kpc$ is the distance to the LMC and $xD$ is the distance
between us and the deflector of mass $M$. In eq.(\ref{einstein})
we also show the value of $R_E$ as a function of $x$ and $M_{0.1}$.
We see that the value of $R_E$ is large compared to the expected
typical radius of the galactic object (for reference $R_{\odot} =
0.0046 AU$).   

Paczy\'nski, in a seminal paper \cite{pac}, evaluated that one
needs to monitor on the order of millions of stars (in the LMC) to
be sensitive to a dark halo in form of machos. The microlensing
signature (when both the source and the lens are single and
point-like) is clear: achromaticity and non-repetitiveness of the
signal, and light-curve magnified by the time-dependent factor

\begin{equation}     \label{A} 
A(t) =  \frac{u^2+2}{u \sqrt{u^2+4}}  \ \ ,
\end{equation}
leading to a magnification with a shape symmetric in time. (The shift
in magnitude is given by $\Delta m = - 2.5 \log A$.) 

In (\ref{A}) the time-dependent parameter $u$ is the distance $d$ of
the macho to the line of sight in units of the Einstein radius

\begin{equation}     \label{u} 
u^2 = \frac{d^2} {R_E^2} = \frac{b^2+(v_T\ t)^2} {R_E^2}  \ \ .
\end{equation}

The distance is of course changing with time since the macho has a
relative transverse velocity $v_T$. The minimum value of the
distance, $b$, is called the impact parameter. It is at the moment
that $d=b$ (closest approach) that the amplification is maximal (it
happens at $t=0$ by definition). 

How large is the maximal magnification depends on the value of
$b/R_E$. When $b$ is less than $R_E$ the change in
the star magnitude at the peak of the effect is more than 0.3 (a
measurable shift). 

As it follows from what we say, $u$ (and $A$) are functions of time;
microlensing is a transient effect that lasts a time on the order of
$T=R_E/v_T$ (see Refs.\cite{pac,derujula1,griest} for details). The
most probable lensing time $T$ depends on the mass of the deflector
and, for a standard halo and velocity distribution, is given by
 
\begin{equation}     \label{time} 
T \simeq 20\ days\ \sqrt{M_{0.1}}  \ \ .
\end{equation}   

Given a microlensing event (and always assuming that the source and
the deflector are point-like), one deduces two parameters: the
maximal amplification and the lensing time $T$. The physical effect,
however, depends on four parameters: $M$, $v_T$, $d$ and the
distance where the lensing took place, $xD$. It follows that in an
event-by-event basis we are not able to identify the precise
values of these four variables.   

In 1990, several groups announced their decision to start
experimental searches. The French collaboration EROS \cite{eros} and
the American-Australian collaboration MACHO \cite{macho,macho2} have
reported the detection of microlensing events, discovered by
long-term photometric monitoring of stars in the LMC. 

When monitoring towards the galactic bulge, a high number of events
have been found, as has been reported by the Polish-American
collaboration OGLE \cite{ogle} and by the MACHO group \cite{macho3}.
These observations can help to further understand the (conventional)
galactic structure, and so may indirectly help to clarify the dark
halo properties. Other collaborations are currently searching for
microlensing: the DUO \cite{duo} and the VATT-Columbia groups. 

The group AGAPE \cite{agape} has started a microlensing search in
the Andromeda galaxy (M31) direction. They monitor pixels rather
than individual M31 stars. The experiment is complementary to other
microlensing searches since the M31 line of sight is very different
from the LMC and bulge directions. As we will see, probing the halo
in various directions will probably be of crucial importance for the
whole issue of brown dwarves in the galactic halo.

The MACHO group has recently presented \cite{macho2} an analysis
of the first year data of LMC results, that correspond to a total
exposure of about $10^7$ star-yr. They find three events consistent
with microlensing. When comparing their results with the expectations
from a dark halo having the form in eq.(\ref{rhoDM}) -they
call this form the standard halo- they are able to assess that the
macho mass fraction $f$ of the dark halo is less than 0.5 in the
macho mass range
 
\begin{equation} \label{machorange} 
3 \times 10^{-4} \leq  \frac{M}{M_{\odot}} \leq 0.06   \ \ .
\end{equation}

Thus for the standard spherical halo (\ref{rhoDM}) machos in the
range (\ref{machorange}) cannot constitute more than half the dark
matter density. However, this limit depends on the halo model, as
emphasized in the MACHO analysis. A very massive halo (with rising
rotation curve) would imply stronger limits, but an extreme
maximal disk model with a very light halo would not give useful
limits \cite{macho2}. A change in the halo shape would also alter the
conclusions. 

A related analysis \cite{gates} takes into account the different
Galactic components: luminous and dark disk, bulge and Galactic
halo with machos and cold DM. They vary parameters as to generate
millions of Galactic models. These models are of course restricted
to agree within the errors with the observational data, including
the microlensing data towards LMC and also towards the bulge. They
conclude that most viable models have $f<0.3$. They find that
Galactic models with a high proportion of machos have a light halo
(this is in agreement with the MACHO group analysis \cite{macho2})
and also distinctive features that could help to test more
definitively whether a (relatively) high proportion of machos in the
halo is excluded or not.     

Apart from the three events found by MACHO, there are two more
events consistent with microlensing that have been detected by 
EROS \cite{eros}. Their second event, if it is indeed microlensing,
does not show the simple form of (\ref{A}); it may correspond to
a binary source star \cite{eros_binary}. These two microlensing
candidates have been detected in 3 years of Schmidt plates data. The
total EROS exposure and efficiency are not drastically different
from the ones from MACHO, and the microlensing times are of the same
order. Thus EROS has similar (but independent) conclusions to the
MACHO ones, in particular the one referring to the range 
(\ref{machorange}).  

In addition, however, EROS has been also performing a CCD
search of short term microlensing events \cite{eros2}, with no
candidates found. The data is sensitive to low-mass brown dwarves
-see eq.(\ref{time})- and the null result has the 
consequence of lowering the tested macho mass range below $10^{-4} 
M_{\odot}$. Testing lower masses is important since below $10^{-4}
M_{\odot}$ there could very well exist compact baryonic DM, since
one is still above the evaporation limit \cite{derujula1} -compare
eqs. (\ref{machorange}) and (\ref{window1}).

In fact, at the time of writing the present review, EROS has
presented a combined analysis of their CCD and photometric data
\cite{eros3}. One of their conclusions is that machos in the range

\begin{equation} \label{erosrange} 
 10^{-7} \leq  \frac{M}{M_{\odot}} \leq 10^{-1}
\end{equation}
cannot contribute more than 50\% to the mass of the standard
galactic halo. Again, the limit (\ref{erosrange}) can be weakened if
the halo is not standard.

We notice that the lower limit in (\ref{erosrange}) is near the
expected cut-off on brown dwarves due to the expected thermal
evaporation,i.e. near $M_{evap}$ in (\ref{evaporation}). In fact
the expected range of DM in (\ref{window1}) and the range
constrained by the EROS result (\ref{erosrange}) coincide. Thus, if
the microlensing bounds are confirmed in the future (we would like
to have a confirmation independent of not-well tested assumptions
on the halo form), the window of brown dwarves in the halo would be
excluded. In this respect, we would like to stress the importance of
testing other directions in the galactic halo, since it
would help to probe its shape and form.     

Let us now address the important question about whether the
detected events are from lensing by objects in our Galactic halo or
in conventional populations like the Galactic or LMC disk, etc. It
turns out that the expected number of events from the known stellar
populations for the first year MACHO data is
$N \sim 0.5$ \cite{gould,macho2}. Thus one may conclude that a new
(dark) Galactic component has been detected.

Unfortunately, as we have already mentioned, given a microlensing
event one cannot deduce from the data at which distance in between
the light source and the observer the event took place. Thus, we do
not know whether this new component is part of the dark Galactic
halo. Still, it is interesting to assume that the microlensing
events are due to objects in the halo, in order to see which are the
consequences.  (To obtain the limits in eqs.(\ref{machorange}) and
(\ref{erosrange}) we do not need to make such an assumption!)

Let us then assume that: 

(A) this new component is the galactic dark halo, i.e., the
detected microlensing events are due to massive objects in the
Milky Way halo. 

(B) the Galactic halo is spherical, and given by eq.(\ref{rhoDM}).

(C) the velocity distribution is maxwellian with a dispersion of
$\sigma \approx  270$ km/s.

With these hypotheses, we can go further and estimate that the lens
masses are most probably in the range \cite{macho2,jetzer}     

\begin{equation} \label{massrange} 
  \frac{M}{M_{\odot}} \simeq 0.04 - 0.06   \ \ .
\end{equation}
although these figures have very large error bars, and that the
fraction $f$ of machos in the halo is
 
\begin{equation} \label{frange} 
f = 0.19^{+0.16}_{-0.10}    \ \ .
\end{equation}
 
To obtain the macho mass function will be a long-term project since
it would require much more data than we have now. The method of mass
moments  \cite{derujula2} could help to reconstruct the mass
function. 

\subsection{Heavy objects}

White dwarves and neutron stars are dark
remnants of stars, and thus they are candidates for baryonic DM. The
problem is that the precursors of the first class have synthesised
and ejected a lot of helium \cite{carr}, and the precursors of the
second class have ejected large quantities of heavy elements
\cite{carr2}. We do not detect these features in the Galactic halo,
and so this makes the two candidates unlikely.       

Above a critical mass $M_{crit}$, supermassive stars undergo complete
collapse to a supermassive black hole and constitute a viable DM
candidate. It is estimated that
 
\begin{equation} \label{crit} 
 M_{crit} \simeq 200 M_{\odot}
\end{equation}

There are upper limits on the mass of supermassive compact objects in
galactic halos. Very heavy objects would disrupt halo globular
clusters \cite{moore} and also would disrupt nearby dwarf galaxies
\cite{rix}, unless

\begin{equation} \label{dis} 
M \leq M_{dis} \simeq 10^{4} M_{\odot} \ \ ,
\end{equation}

The allowed range of masses for supermassive black holes is between
the bounds (\ref{crit}) and (\ref{dis}). This a second mass window on
compact baryonic DM:

\begin{equation} \label{window2} 
10^2 M_{\odot} \simeq M_{crit}   \leq  M \leq 
 M_{dis} \simeq 10^4 M_{\odot} \ \ .
\end{equation}

\section{CONCLUSIONS}

We have compelling evidence for galactic and baryonic DM. 
It is important to stress that the need for galactic DM
is based on
\vskip 0.3truecm 
{\bf (1)} the observed galactic rotation curves  

{\bf (2)} newtonian mechanics.
\vskip 0.3truecm
The experimental search for galactic DM is thus independent of
theoretical conjectures as inflation, etc.

The need for baryonic DM is based on 
\vskip 0.3truecm 
{\bf (a)} the theory of BB nucleosynthesis  

{\bf (b)} the observation of the yields of light elements
and the deconvolution of the chemical evolution to deduce the
primordial abundances. 
\vskip 0.3truecm 

The natural place for dark baryons is in galaxies, and thus most
probably they constitute at least a fraction of the dark galactic
halos. These galactic dark baryons are probably not in diffuse form
(although such a possibility is not completely excluded). The
possibility that they are non-shining light objects is restricted by
a variety of arguments, as self-destruction by collisions, Earth
hitting and thermal evaporation. 
Star remnants (white dwarfs and neutron stars) have contaminated the 
Galaxy beyond the observed yields of elements and thus are unlikely 
candidates. 

Compact baryonic objects could be very heavy since then there is no
ejection of matter in the star history. However, too heavy objects
would disrupt astronomical systems and they are also excluded.

Two windows for $M$ remain, the one that corresponds to brown 
dwarves (\ref{window1}), and the one that corresponds to 
supermassive black holes (\ref{window2}). 

It is remarkable that the first window can be tested with
microlensing experiments. One has to keep in mind however
that this effect is independent of the constituency of the
compact object. The second window, although there is 
no compelling scenario where such supermassive black holes are 
produced, has to be considered as candidate for a dark halo of
baryonic constituency.

\vskip 0.3truecm

{\bf Acknowledgements}. I would like to thank Angel Morales and
other organizers of the TAUP 95 workshop for the nice atmosphere they
helped to create at the workshop. I thank the
Theoretical Astroparticle Network for support under the EEC Contract
No. CHRX-CT93-0120, and also the Spanish CICYT Research Project No.
AE-93-0520.

\end{document}